\documentclass[twocolumn,preprintnumbers,nofootinbib,prl]{revtex4}

\usepackage{graphicx}

\usepackage[hypertex]{hyperref}
\newcommand{\beq}{\begin{equation}}
\newcommand{\eeq}{\end{equation}}
\newcommand{\be}{B_\oplus}

\def\be{\begin{equation}}
\def\ee{\end{equation}}
\def\baray{\begin{eqnarray}}
\def\earay{\end{eqnarray}}
\def\ba{\begin{eqnarray}}
\def\ea{\end{eqnarray}}

\begin{document}

\pagestyle{plain}

\preprint{UCB-PTH-07/08, MADPH-07-1491, MAD-TH-07-08}

\title{Probing the Geometry of Warped String Compactifications at the LHC}

\author{Gary Shiu}
\affiliation{Department of Physics, University of Wisconsin,
Madison, WI 53706, USA}
\author{Bret Underwood}
\affiliation{Department of Physics, University of Wisconsin,
Madison, WI 53706, USA}
\author{Devin G.E. Walker}
\affiliation{Department of Physics, University of California,
Berkeley, CA 94720, USA}
\affiliation{Theoretical Physics Group, Lawrence Berkeley National Laboratory,
Berkeley, CA 94720, USA}
\author{Kathryn M. Zurek}
\affiliation{Department of Physics, University of Wisconsin,
Madison, WI 53706, USA}


\begin{abstract}

Warped string compactifications, characterized by non-singular behavior of the metric in the infrared (IR), feature departures from the usual anti-de Sitter warped extra dimensions. We study the implications of the smooth IR cutoff for Randall-Sundrum (RS) type models.
We find that the phenomenology of the KK gravitons (including their masses and couplings) depends sensitively on the precise shape of the warp factor in the IR.  In particular, we analyze the warped deformed conifold and find that the spectrum differs significantly from that of RS, and present a simple prescription (a mass gap ansatz) which can be used to study 
the phenomenology of IR modifications to
 5-d warped extra dimensions.

\end{abstract}
\maketitle

Theories with more than three spatial dimensions occur recurrently
in attempts of unification: from the original idea of Kaluza and Klein to its modern manifestation in 
string theory. 
Extra dimensions also emerge in extensions of the Standard Model,
most notably, in bottom-up approaches \cite{Arkani-Hamed:1998rs,RS}
to solve the hierarchy and flavor problems.
Since extra dimensions appear ubiquitously in both top-down and bottom-up 
studies,
it is important to investigate how we can distinguish between different extra dimension scenarios and to what extent we can learn about the nature of the underlying theory from the size and shape of the extra dimensions.

If the extra dimensions are large \cite{Arkani-Hamed:1998rs} (around mm lengths), then the signatures can be quite dramatic, such as missing energy at colliders or
deviations from the inverse square law at small, but measurable, distances \cite{Adelberger:2003zx}.
 If the dimensions are small, then detection becomes much more difficult; the high energies associated with such small length scales are 
hard
 to access, though
it has recently been argued that
details of their shapes may leave 
imprints on the Cosmic Microwave Background 
via
inflation 
\cite{DBIObs}.
 As shown by Randall and Sundrum \cite{RS}, however, small extra dimensions may be observed at colliders if the extra dimensions are strongly warped.  The signature is the appearance of TeV mass spin-2 Kaluza-Klein (KK) gravitons which decay to pairs of Standard Model particles 
\cite{OriginalWarpedGravitons, WarpedGravitons}.  TeV scale masses arise from an
exponentially suppressed Planck scale, $m \sim M_{pl} e^{-k \pi r_c}$, generated by the spatial warping.  Extensions of the original RS model 
may also 
solve the flavor structure of the Standard Model (SM) in a geometrical way by localizing the SM fields at different
locations in the warped extra dimension \cite{RSFlavor}.

Warped extra dimensions 
arise in string theory through a somewhat different route.
In recent attempts to stabilize moduli in string theory, it was noted that compactifications with fluxes 
\cite{Douglas:2006es} 
can result in strongly warped spaces, sometimes known as warped throats.
While many such string constructions contain asymptotically an Anti-deSitter (AdS)
region, their warped geometries may differ from AdS in the infrared (IR). 
This feature can be understood from the gauge/gravity correspondence: the field theory dual of such warped throats may not be conformal, in which case the warp factor receives log corrections, and the smooth IR behavior is associated with the existence of a strong dynamics scale.
An example of such warped throats is the warped deformed conifold \cite{KT,KS} where 
the warp factor approaches an exponentially small constant near the tip.
 In this
 Letter
  we 
  discuss
  how the detailed geometry of the warped extra dimension dramatically changes the phenomenology of RS type models.  Warped extra dimension model building has been analyzed almost exclusively within the context of AdS$_5$.  Here we
 compare and contrast RS models to
  warped string compactifications with a smooth IR cutoff.
  We find that RS type phenomenology is significantly changed by these modifications to the AdS metric; in particular, 
   the spacing between graviton KK modes and the strength of their couplings
   are strongly altered. 
Such strong differences offer 
the tantalizing possibility
of distinguishing
different 
warped geometries experimentally.
Thus top-down string constructions could potentially 
have direct impact for observations at the LHC.

In type IIB flux compactifications of string
theory \cite{GKP}, the metric takes the generalized form
\begin{equation}
ds^2_{10} = f^{-1/2}(r)\hat{g}_{\mu\nu}dx^{\mu}dx^{\nu} + f^{1/2}(r) (dr^2 + ds_{X^5}^2)\, .
\label{eq:WarpedMetric}
\end{equation}
For a warp factor
$f(r) = \frac{R^4}{r^4}$ the first two terms in the metric are easily recognized as the 5-d RS metric with the identification $e^{-k y} = r/R$, while the last term introduces a new 5-d angular space $X^5$.  The hierarchy between the electroweak and UV scales arises in the same way as in RS, as a ratio between the warp factor at the UV and IR branes, $\left(f(r_{IR})/f(R)\right)^{-1/4} = \Lambda_{IR}/\Lambda_{UV}\equiv r_{tip}$,  where the IR brane is said to reside at the ``tip'' of the 
throat.

\begin{figure}[t]
\begin{center}
\includegraphics[scale=.4]{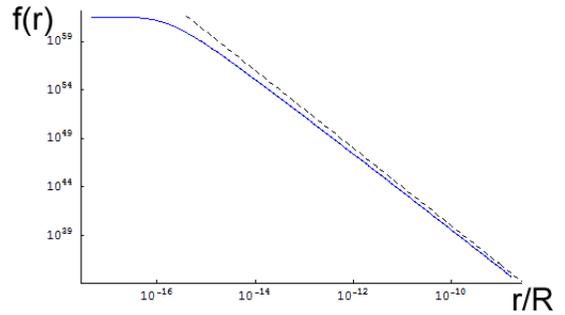}
\caption{KS (solid blue) and RS (dashed black)
warp factors $f(r)$ as a function of the radial coordinate $r$.
}
\label{fig:RSKSPlots}
\end{center}
\end{figure}

While RS is 
motivated
 from a low-energy effective theory point of view for solving the hierarchy problem, 
 warped geometries from flux compactifications in string theory may not be exactly AdS.
 An example is 
 a
  solution in type IIB supergravity 
 known as the warped deformed conifold, or the so called
Klebanov-Strassler (KS) throat.
Far from the IR end of the throat, the KS warp factor
mimics
the RS solution, where the geometry
is approximately given by the warped product
$\mbox{AdS}_5\times T^{1,1}$, and
$T^{1,1}$ is a 5-dimensional Einstein-Sasaki space with topology $S^2\times S^3$.  In RS, the warp factor displays singular behavior as $r \rightarrow 0$; this is resolved by truncating the space and placing the IR brane at non-zero $r$.  In KS, however, the singular behavior of the RS geometry is smoothed out and the 
KS warp factor approaches a constant as $r \rightarrow 0$.  The angular modes are integrated out, and what remains is a complicated function of $r$, which 
we write as $f(r) = (R/r_{tip})^{4} I(\tau)/I(0),$ where $\tau = \tau(r)$.  We plot in fig.~\ref{fig:RSKSPlots} the KS and AdS warp factors $f(r)$ as a function of the 
radial coordinate $r$.  Notice that the KS solution is
well-approximated by the AdS geometry for much of its length, with small differences showing up only very near the IR end of the
throat.  Due to the different behavior of KS and RS as $r \rightarrow 0$, the weak-Planck hierarchy results from the warp factor evaluated at different points:  $r=\tau = 0$ for KS and $r = r_{tip}$ for RS.

Since the 
gravitational KK modes 
are 
peaked near the IR brane, as in RS, we expect these differences in the warp factor to
show up as differences in the KK spectra.  To determine how significant these differences are, we solve the equation of motion derived from the 10-d Laplacian on the warped space, for a $4-$d gravitational KK mode $\hat{g}_{\mu\nu} = \eta_{\mu\nu} + h_{\mu\nu}(x,r)$ 
with the decomposition $h_{\mu\nu}(x,r) = \overline{M}_p^{-1} \sum_n h_{\mu\nu}^{(n)}(x) \phi_n(r)$,
\begin{equation}
\partial_r\left(\sqrt{g_{X^5}} \partial_r\phi_n(r) \right) + m_n^2 \sqrt{g_{X^5}} f(r) \phi_n(r) = 0\, .
\label{eq:KKEOM}
\end{equation}
Here $g_{X^5}$ is the determinant of the $5-$d space $X^5$ and we defined the KK mass $m_n$ by 
$\hat{\nabla}_4 h_{\mu\nu}^{(n)}(x) = m_n^2 h_{\mu\nu}^{(n)}(x)$.  
(The KK mass spectrum of   
the KS throat 
was
previously considered
in \cite{HH, Berg:2006xy}, 
while 
the KK modes of \cite{Yamaguchi} 
do not satisfy 
Eq.~(\ref{eq:KKEOM}). 
Here we consider the full problem of both the masses {\it and} the couplings obtained from (\ref{eq:KKEOM}).)

\begin{figure}[t]
\begin{center}
\includegraphics[scale=.4]{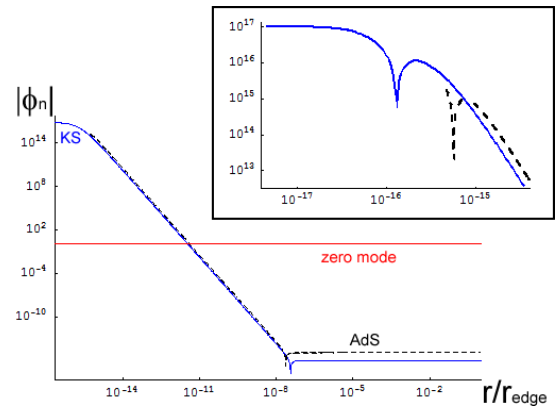}
\caption{Normalized KK graviton wave functions for the zero mode (solid red) and for the first KK mode of the AdS (dashed black)
and KS (solid blue) geometries.  Inset:
The second  KK modes for RS (dashed black) and KS (solid blue).}
\label{fig:WaveFunctionPlots}
\end{center}
\end{figure}

Interestingly, the equation of motion (\ref{eq:KKEOM}) for $10-$d KK modes on $\mbox{AdS}_5\times S^5$ is the same as for
$5-$d KK modes on just $\mbox{AdS}_5$, so the KK mode solutions will be identical in both cases.  The KS equation of motion
can be solved by requiring that the KK wavefunction be finite at the IR end of the throat and asymptotically approach the AdS solution
in the UV.  
We plot the results in fig.~\ref{fig:WaveFunctionPlots} for RS and KS metrics.  There are two notable 
features.  
First, the KS wavefunction has a larger profile on the IR brane than does RS (which are located in different places anyhow, with the IR brane located at $r=0$ in KS, but at $r = r_{tip}$ in RS).  While RS truncates at $r = r_{tip}$, the KS solution continues to grow and then level out at $r \ll r_{tip}$.  
The second feature is that the higher KK modes in KS have a larger profile in the IR relative than the lower KK modes,
which
implies stronger couplings to IR brane confined fields.  This is a simple result of the oscillation features of the higher KK modes; the KK wavefunction oscillates to a minimum just as the 
Jacobian $\sqrt{\tilde{g}}$ (the determinant of the 
6-dimensional 
metric $ds^2_6 = dr^2+ds^2_{X^5}$) weighting the normalization condition is peaking in $r$.  As a result, the overall normalization of these higher modes must be higher than for the lower KS KK modes, resulting in a larger overlap on the IR brane.

The different profiles of the KK gravitons on the brane and in the bulk have interesting phenomenological consequences through their couplings to 
SM
fields.  The KK modes couple to matter through the stress-energy tensor,
\begin{equation}
{\mathcal L} = \frac{1}{\overline{M}_p} h_{\mu\nu}^0 T^{\mu\nu} + \sum_{n>0} \frac{1}{\Lambda_{KK}^{(n)}}h_{\mu\nu}^n T^{\mu\nu}
\label{eq:KKMatterCoupling}
\end{equation}
where 
$\Lambda_{KK}^{(n)} \equiv \overline{M}_p/ \phi_n(r_*)$
for matter confined on the IR brane at $r_*$.  For matter with a 
profile in the bulk the coupling is straightforwardly 
generalized to the overlap integral of the KK wavefunction with the matter profile.

The KK graviton masses and couplings to SM fields confined to the IR brane are shown in Table~\ref{tab:RSKSTable}, where we have fixed the total warping to $r_{tip} \sim 2 \times 10^{-14}$ (as opposed to 
 $r_{tip} \sim e^{- 11.27 \pi} \sim 4.2 \times 10^{-16}$ usually chosen to solve the hierarchy problem).  The differences between KS and RS are significant. The KS masses are more
finely spaced than RS (with dimensionless separations of $(m_i-m_{i-1})/m_1 \sim 0.51, 0.83$ respectively) by a factor of $\sim 1.6$.  
The more dramatic differences reside in the couplings
-- the larger profile on the IR brane of the KS modes implies much stronger couplings.  The couplings are so strong that if photons and electrons reside on the IR brane and the usual warp factor $r_{tip} \sim 4.2 \times 10^{-16}$  to solve the hierarchy is chosen, a KS warped extra dimension would already have been observed at the Tevatron \cite{Abazov:2005pi,Abulencia:2005nf}.  Another interesting feature is that, while the couplings are universal in RS, the couplings are non-universal in KS and become more strongly coupled with higher KK mode, as we already noted on account of the higher overlap on the IR brane.  We see 
here 
then that even though the KS warped throat is well-approximated
by an AdS geometry for most of its length, small differences in the IR region lead to 
vastly different masses of the KK 
modes
and their couplings to the 
SM.

We are interested in studying a larger class of warped supergravity solutions, 
e.g., warped throats in \cite{BGMPZ}
which are related to KS, having different
angular isometries and 
 IR behavior of the warp factor.
Unfortunately the full solutions can only be evaluated
numerically.  It is useful, therefore, to parameterize a large class of
warp factors with a ``mass gap" ansatz \cite{DBITip} for the purpose of exploring the effects of the metric on warped phenomenology
\begin{equation}
f(r) = \frac{R^4}{r_{tip}^4 + f_2 r^2 + r^4},
\label{eq:MassGap}
\end{equation}
where the coefficients of $r_{tip}^4$ and $r^4$ terms are fixed respectively by the hierarchy and AdS behavior at large $r$.  By 
varying
 the coefficient $f_2$
we can interpolate between different warp factors.  Note that the ``mass gap" ansatz can
also be used to construct warp factors with smooth IR behavior in $5$ dimensions, where the metric takes the form $ds^2 = f^{-1/2}(r)\hat{g}_{\mu\nu}dx^{\mu}dx^{\nu} + f^{1/2}(r) dr^2$.
These metrics can  be considered as a smooth extension of the RS scenario in the IR without
the need for an explicit cutoff.
They may offer new model building possibilities for RS
scenarios, which we hope to return to in a future work.

Although small changes in the coefficient $f_2$ of the mass-gap metric leads to small changes in the behavior of the warp factor
at the IR end of the throat, these small changes can lead to large differences in the spectrum of the
masses and couplings of the KK modes.    The spectrum is shown in table~\ref{tab:5dMassGap} for different values
of the mass gap parameter $f_2$ in $5-$d with the standard 
weak-Planck hierarchy
$r_{tip} = 4.2 \times 10^{-16}$; 
the differences
in the $10-$d spectra as $f_2$ is changed are qualitatively similar.
(Differences in the KK spectra due to small changes in the geometry were also considered in
the context of toroidal compactification in \cite{TorusShape}).

\begin{table}
\begin{tabular}{|c|cc|cc|} \hline
KK Mode & RS Mass & KS Mass & RS Coupling & KS Coupling \\ \hline 
1 & 1.000 & 1.000 & 51.15 & 1.975 \\ \hline
2 & 1.831 & 1.506 & 51.15 & 1.244 \\ \hline
3 & 2.655 & 2.012 & 51.15 & 0.921 \\ \hline
4 & 3.477 & 2.519 & 51.15 & 0.737 \\ \hline
5 & 4.298 & 3.027 & 51.15 & 0.619\\ \hline
\end{tabular}
\caption{RS and KS Masses and Couplings ($\Lambda_{KK}$ in TeV) with warp factor $r_{tip} \sim 2 \times 10^{-14}$.}
\label{tab:RSKSTable}
\end{table}

\begin{table}
\begin{tabular}{|c|cc|cc|} \hline
KK Mode & $f_2=0$ & $f_2 = 2 r_{tip}^{1.9}$ & $f_2=0$ & $f_2 = 2 r_{tip}^{1.9}$ \\ 
 & Mass & Mass & Coupling & Coupling \\ \hline 
1 & 1.000 & 1.000 & 1.11 & 0.806 \\ \hline
2 & 2.410 & 1.480 & 1.35 & 0.765 \\ \hline
3 & 3.596 & 2.086 & 1.38 & 0.796 \\ \hline
4 & 4.756 & 2.717 & 1.38 & 0.814 \\ \hline
5 & 5.907 & 3.354 & 1.39 & 0.823 \\ \hline
\end{tabular}
\caption{Masses and Couplings for $5-$d Mass Gap ($\Lambda_{KK}$ in TeV) with warp factor $r_{tip} \sim 4.2 \times 10^{-16}$.}
\label{tab:5dMassGap}
\end{table}

\begin{figure}
{
\includegraphics[width=8truecm,height=7truecm,clip=true]{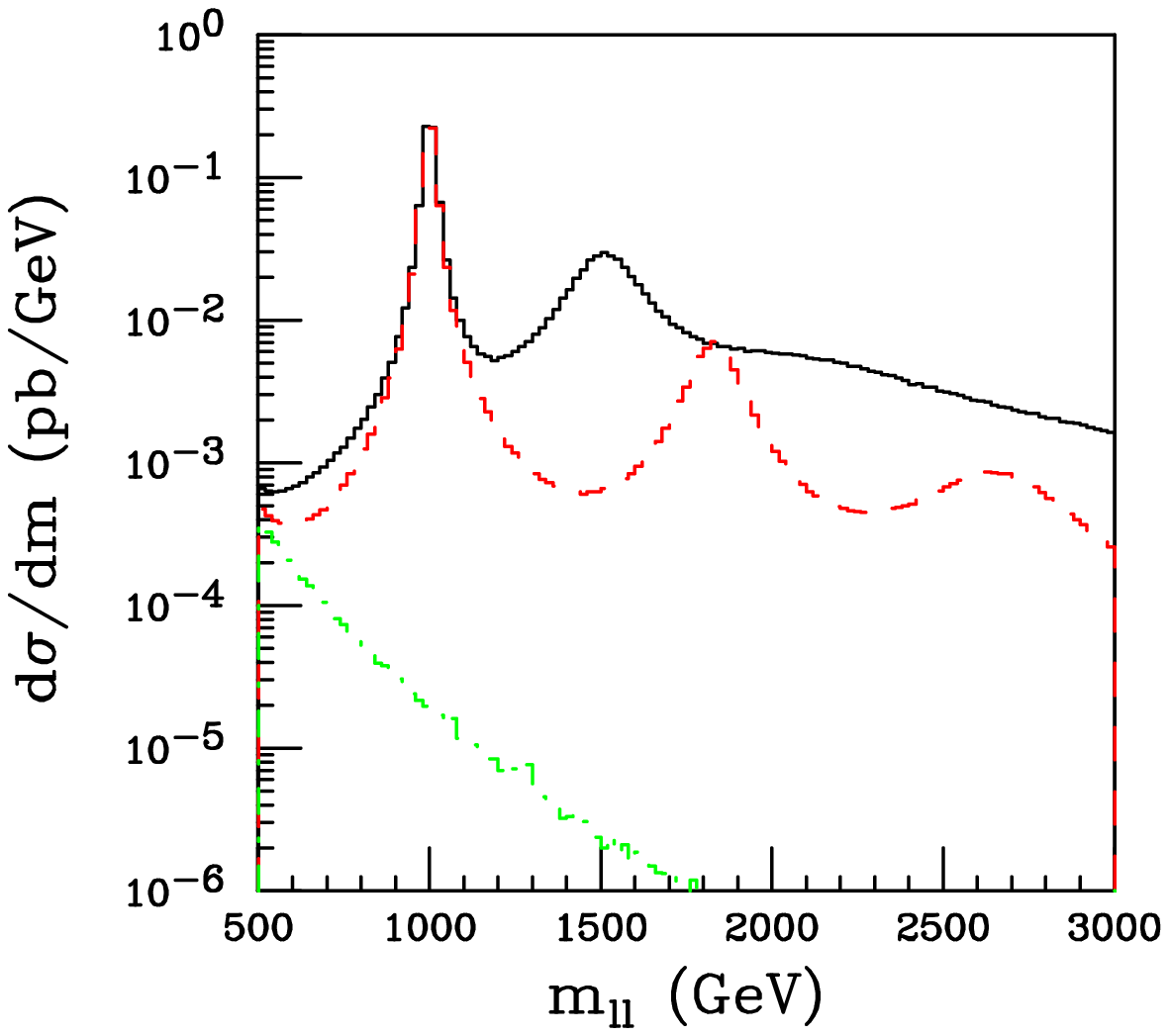}}
\vspace{0.75cm}
{	
\includegraphics[width=8truecm,height=7truecm,clip=true]{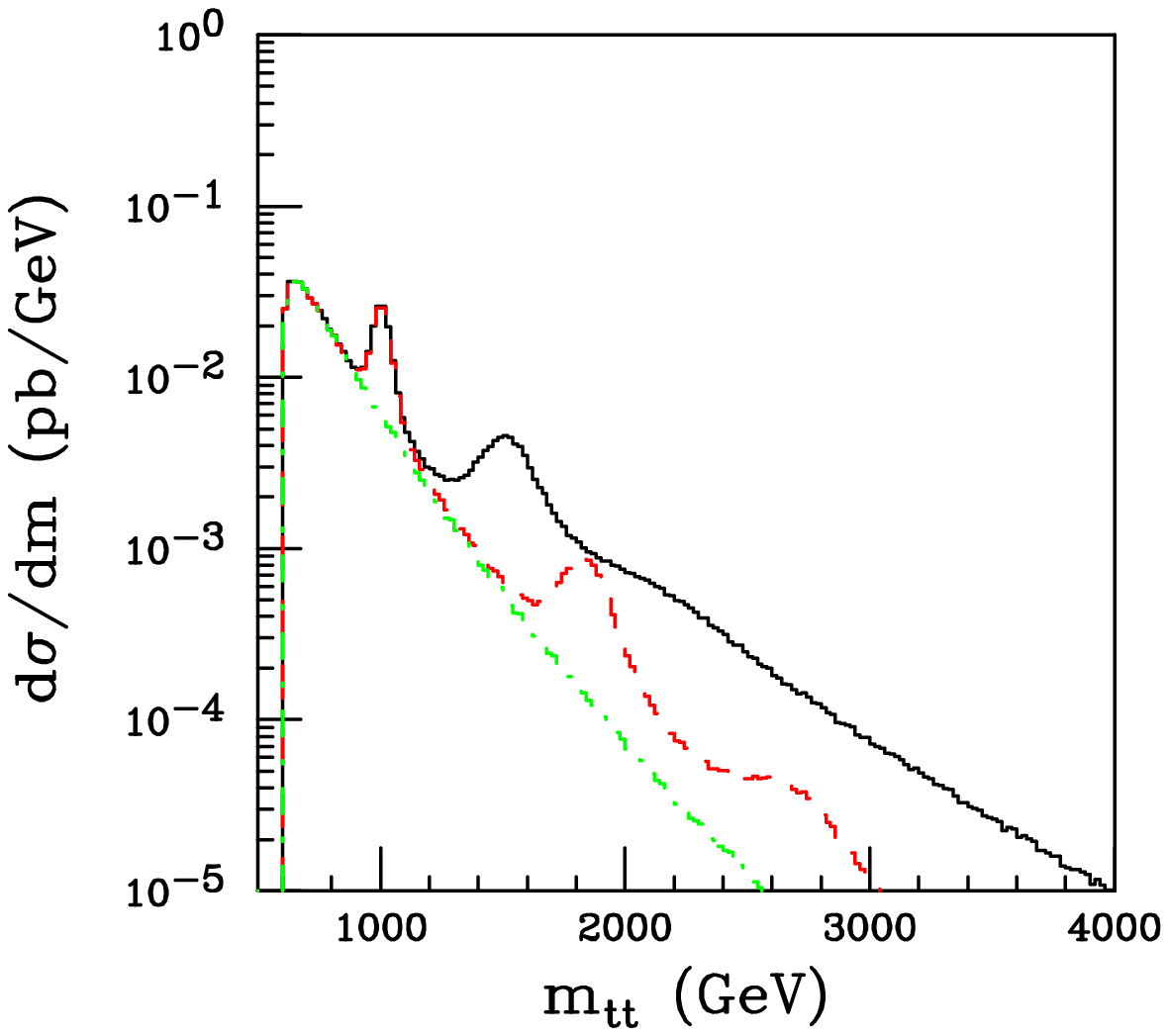}}
\caption{$\ell^+ \ell^-$ (above) and $t \bar{t}$ (below) invariant mass distributions for KS (black solid) with hierarchy $r_{tip} = 2 \times 10^{-14}$ (masses and couplings given in Table I)
and RS (dashed red) with hierarchy $r_{tip} = 8 \times 10^{-16}$ (masses given in Table I,
with $\Lambda_{KK}=2$ TeV).  The SM backgound (dot-dashed green) is $Z,\gamma \to \ell^+ \ell^-$ and gluon $\to t \bar{t}$, respectively. In generating these plots, we have assumed that all Standard Model fields are confined to the IR brane.}
\label{Fig:plots}
\end{figure}

Details of the warped KK graviton spectra such as the spacing of the modes and the strength and pattern of their couplings, which
depend sensitively on the specifics of the IR geometry, can appear as interesting signatures in resonant
production of KK modes at the LHC.  
We will consider the resonant production of a KK graviton from quark and gluon fusion and decay via dileptons 
$q\bar{q}, gg \rightarrow h_{KK} \rightarrow \ell^+\ell^-$ or a pair of tops 
$q\bar{q}, gg \rightarrow h_{KK} \rightarrow t\bar{t}$.  The former channel is much cleaner, but will only become comparable 
when light fermions are localized near the IR brane.  The latter process is more important in models which solve the flavor problem through localizing the light fermions toward the Planck brane. 
The cross section and decay width can easily be worked out from \cite{Tao}:
we show in fig.~\ref{Fig:plots} the differential cross section
for resonant production of KK gravitons for the KS and RS geometries for both the dilepton and top 
semi-leptonic
decay channels.  
The choices for detector
and acceptance cuts, as well as, a method to verify the spin of the KK
gravitons can be found in ref.~\cite{Tops}.  The plots are made in consideration
of $100 fb^{-1}$ of data at design luminosity.
Here we compare KS and RS for two different choices of the warp factor: $r_{tip} \sim 2 \times 10^{-14}$ for KS and $r_{tip} \sim 8 \times 10^{-16}$ for RS.  The latter choice is more standard for solving the hierarchy problem.  The warp factor $r_{tip} \sim 2 \times 10^{-14}$ was taken for KS since $r_{tip} \sim 8 \times 10^{-16}$ results in such large couplings that KS gravitons would have already been observed.  
 The somewhat milder KS warping may in fact be 
motivated since the 
 UV scale (e.g., string scale) could be 
 lower than the Planck scale.
 There are two features of the KS versus RS spectra which are independent of the choice of the warp factor, and can be observed in fig.~\ref{Fig:plots}.  First, KS modes are more closely spaced than RS modes (the size of the dimensionless spacing, $(m_i -m_1)/m_1$, is independent of the warp factor).  The presence of two closely spaced KK resonances is a distinctive feature of IR modifications of the AdS metric
 from 
  string compactifications.  Second, since KS couplings become stronger at higher KK mode, whereas RS couplings are universal, KS KK modes quickly become very broad, and contribute to the signal only through an overall excess of events over the background.  The observability of these effects, of course, depends on the overall coupling $\Lambda_{KK}$, which is determined not only by the total warp factor generating the hierarchy, but now also by the background geometry.      

In summary, 
because the
KK modes are 
peaked at the IR end of a warped throat, small differences in the 
IR geometry 
can significantly modify the details of the KK graviton 
spectrum and their observable signatures
even when the warp factors of these geometries differ only slightly.  
While we explicitly demonstrated such
differences by comparing the RS model with the KS throat,
it would be interesting to extend our analysis to other string theory warped throats with different isometries and IR behavior. 
Importantly, details of the warped geometries may also
leave imprints on 
cosmological data
 \cite{DBIObs}. 
Hence, combined data from particle physics and cosmology may help 
us uncover the nature of extra dimensions
and zero-in on a set of promising string vacua.

We thank I.~Hinchliffe, M.~Shapiro, H.~Tye for discussions.
G.S., B.U. and K.Z. are supported in part by NSF CAREER Award No. PHY-0348093, DOE grant DE-FG-02-95ER40896, a Research Innovation Award and a Cottrell Scholar Award from Research Corporation. D.W. is supported by
a University of California Presidential Fellowship.

\end{document}